\documentstyle[12pt]{article}

\newcommand{\beq}{\begin{equation}}
\newcommand{\eeq}{\end{equation}}

\newcommand{\beqcol}{\begin{array}{rcl}}
\newcommand{\eeqcol}{\end{array}}

\newcommand{\mattwo}[4]{\left( \begin{array}{cc}
			 #1    &   #2   \\
			 #3    &   #4   \\       \end{array} \right)}

\newcommand{\mattre}[9]{\left( \begin{array}{ccc}
			#1  &  #2  &  #3 \\
			#4  &  #5  &  #6 \\
			#7  &  #8  &  #9  \\      \end{array} \right)}

\newcommand{\indf}[2]{^{#1}_{\;\:\, #2}}
\newcommand{\indfd}[2]{^{\;\;\, #1}_{#2}}
\newcommand{\indcd}[2]{^{\tilde{#1}}_{\;\;\, \tilde{#2}}}
\newcommand{\indc}[2]{^{\;\;\,\tilde{#1}}_{\tilde{#2}}}

\newcommand{\Pxi}[1]{\left( P_{\xi_{(#1)}} \right)}
\newcommand{\Pxib}[1]{\left( P_{\bar{\xi}_{(#1)}} \right)}

\newcommand{\C}{\bf{C}}

\newcommand{\cld}{C-q\lambda D}

\newcommand{\lam}{\lambda}
\newcommand{\alf}{\alpha}
\newcommand{\eps}{\varepsilon}

\newcommand{\suq}{SU_q(2)}
\newcommand{\slq}{SL_q(2,\C)}

\newcommand{\xihf}{\xi_{({1\over 2})}}
\newcommand{\half}{{1\over 2}}

  \newcommand{\ket}[1]{\left| #1 \right\rangle}

\newcommand{\scpro}[2]{\left\langle #1\mid #2\right\rangle}

\newcommand{\tde}[1]{\tilde{#1}}

\newcommand{\binom}[2]{ \left[ \begin{array}{c} #1 \\ #2
                               \end{array} \right]_{q^{-2}}^{\half} }

\begin{document}


{\thispagestyle{empty}
\begin{flushright} MPI-Ph/93-61 \\ July 1993 \end{flushright}
\vfill
\begin{center}
{\Large
 $q$-Deformed Relativistic Wave Equations
}
\\
\vspace{2cm}
{\sc Mathias Pillin}\footnote{Supported by Studienstiftung des deutschen
Volkes}  \\
\vspace{1cm}
Max-Planck-Institut f\"ur Physik \\
Werner-Heisenberg-Institut       \\
F\"ohringer Ring 6               \\
 D-80805 M\"unchen, Germany       \\

\medskip

\vfill
{\bf Abstract}
\end{center}
\begin{quote}
Based on the representation theory of the $q$-deformed Lorentz and
Poincar\'e symmeties $q$-deformed relativistic wave equation are
constructed. The most important cases of the Dirac-, Proca-,
Rarita-Schwinger- and Maxwell- equations are treated explicitly. The
$q$-deformed wave operators look structurally like the undeformed ones but
they consist of the generators of a non-commu\-ta\-tive Minkowski space. The
existence of the $q$-deformed wave equations together with previous
results on the representation theory of the $q$-deformed Poincar\'e
symmetry solve the $q$-deformed relativistic one particle problem.
\end{quote}
\eject
}

\setcounter{page}{1}
\section{Introduction}

\bigskip

In recent years quantum groups have been studied extensively by
both mathematicians and physicists. A special point of interest is
the quantum deformation of the symmetries of flat space-time,
the Lorentz and Poincar\'e groups \cite{CSSW,PoWo,LOR,POI,LUK}.

Following the ideas of Wigner \cite{WIG}, uni\-tary ir\-re\-duci\-ble
re\-pre\-sen\-ta\-tions
(Irreps) of the $q$-deformed Poincar\'e symmetry \cite{POI} have
been investigated to construct massive \cite{PSW} and massless \cite{TALK}
representations of that symmetry. These Irreps can be interpreted as
$q$-deformed relativistic one-particle states. It turns out that these
states have a number of interesting properties. For example the
momentum spectrum is discrete and the mass becomes quantized in terms
of the deformation parameter $q$.

The construction of \cite{PSW} only led to spinless massive one-particle
states. To incorporate the spin, covariant- or spinor-bases for the
$q$-deformed Poincar\'e algebra have been invented in \cite{PW} using
finite dimensional corepresentations of the Quantum Lorentz Group (QLGr)
\cite{PoWo}. As in the undeformed case these spinor bases carry by
construction more spin degrees of freedom than the physical particle which
they ought to describe. The additional degrees of freedom can be
removed classically by covariant subsidiary conditions which turn out to be
the relativistic wave equations. These subsidiary conditions can be obtained
using only techniques of group representation theory \cite{NoRAF}.

The aim of this work is to show that $q$-analogues of relativistic
wave equations can be constructed with the help of representation
theory of quantum groups. The $q$-deformed wave equations then have
the same properties as the undeformed ones. It should be noted that
a $q$-analogue of the Dirac equation has already been constructed
following a different approach \cite{ARNE}.

The outline of the paper is as follows. In section 2 the
classical construction of relativistic wave equations based on
representation theory is briefly explained. Section 3 and 4 review the
necessary material of the representation theory of the QLGr and the
$q$-deformed Poincar\'e symmetry. In the following sections the $q$-deformed
Dirac-, spin 1 Joos-Weinberg-, Maxwell-, Proca- and Rarita-Schwinger-
equations are constructed. In section 9 the general case of a $q$-deformed
relativistic wave equation for arbitrary spin is considered.

\section{The classical situation}

\bigskip

In order to make the construction of the $q$-deformed wave equations
more transparent the procedure in the classical case is briefly outlined.
Only the massive case is treated here. More details can be found in
\cite{NoRAF}. The construction is based on the representation theory of
semidirect products using the method of induced representations.

When working with Mackey- or Wigner- states the unitary Irreps of the
Poincar\'e group can be induced directly by a given unitary Irrep
$D(k)$ of the stability subgroup $SU(2)$. It is possible to find a special
transformation from the
Mackey- or Wigner- states to the so called covariant- (spinor-) states
for the Poincar\'e group. From the viewpoint of the inducing procedure
this means that $D(k)$ is imbeded into a representation
${\cal D}^{(m,n)}(g)$ of the Lorentz group whose restriction to $SU(2)$,
${\cal D}(k)$, is unitary. $m$ and $n$ denote the highest weights of
the Weyl representation of the Lorentz group. An important property of
the spinor states is that the inner product of two states $\psi_1$
and $\psi_2$ is given by:

\beq
\left(\psi_1,\psi_2 \right) = \int \mbox{d}\mu(p)
  \left(\psi_1(p), {\cal D}^{\dagger}(s){\cal D}(s) \psi_2(p) \right)_{\cal D}
\label{ipr1}\eeq

The variable $p$ denotes the momenta by which the states are diagonalized,
d$\mu(p)$ is some measure in momentum space and ${\cal D}(s)$ is some
finite dimensional representation of $s \in SL(2,\C)$.

The problem is that ${\cal D}(k)$ contains in general other representations
of $SU(2)$ than the desired one $D(k)$. Hence the unrequired representations
occuring in ${\cal D}(k)$ have to be eliminated covariantly. This makes
covariant subsidiary conditions necessary, which are the wave equations.

If the momentum eigenvalue $p$ occuring in the arguments of the states in
(\ref{ipr1}) is taken to be obtained by a pure Lorentz boost from a
rest system vector $p_r$: $ p = {\cal D}^{(1/2,1/2)}(s) p_r$, the expression
(\ref{ipr1}) can be simplified to:

\beq
\left(\psi_1,\psi_2\right) = \int {{{\mbox d}^3 p}\over{\omega}}
                       \psi_1^{\dagger}\;{\cal D}^{-1}
                 \left({{\sigma.p}\over {m}}\right) \psi_2(p),
 \label{ipr2}
\eeq
where $m$ is the mass eigenvalue and $\sigma$ represents the ordinary Pauli
matrices.

A wave operator ${\cal W}^{(j)}(p)$ is a projection operator from the set of
$SU(2)$ representations in ${\cal D}(k)$ to the desired spin $j$
representation $D^{(j)}(k)$. It can be shown that
this wave operator is in principle equivalent to the object
${\cal D}^{-1}\left( {{\sigma.p}\over{m}} \right)$ in (\ref{ipr2}). To
establish the
equivalence completely certain index symmetry properties have to be taken
into account when the wave operator is applied to a wave function.

The following important theorem holds in general for in\-duced
re\-pre\-sen\-tations:

\medskip

{\bf Theorem}: \quad The covariant form of an induced representation is
completely characterized by the set $\{ {\cal D}(g), {\cal W}\}$
consisting of an inducing representation ${\cal D}(g)$ of the full
Lie group and a projection operator ${\cal W}$ to the required unitary
representation $D(k)$ of the stability subgroup.

\bigskip

In the special case of the Poincar\'e group this means that a massive
spinor basis consisting of a direct product of a spinless unitary Irrep
and a finite-dimensional representation of the Lorentz group together
with a covariant wave operator which restricts the spin degrees of freedom
to the physical ones in all Lorentz frames completely solves the massive
relativistic one particle problem in momentum space.

\section{The Quantum Lorentz Group}

\bigskip

The Quantum Lorentz Group (QLGr) can be defined by considering two
copies of a deformed $SU(2)$ Hopf algebra corresponding to the non-equivalent
fundamental representations of the quantum group $\slq$. We take
the deformation parameter $q > 1$, and the abbreviation
$\lambda = q-q^{-1}$ is sometimes used.

One starts by introducing a quantum matrix $M\indf{\alf}{\beta} :=
\mattwo{a}{b}{c}{d} \indf{\alf}{\beta} $. The entries of
$M\indf{\alf}{\beta}$ obey commutation relations which are generated by
the R-matrix \cite{CSSW} of $SL_q(2)$:

\beq
 \hat{R}^{\alf\beta}{}{}_{\gamma\delta}
  M\indf{\gamma}{\mu} M\indf{\delta}{\nu} \; = \;
 M\indf{\alf}{\rho} M\indf{\beta}{\sigma}
 \hat{R}^{\rho\sigma}{}{}_{\mu\nu}
\label{RTT1}
\eeq

The Hopf algebra structure of the quantum matrix $M\indf{\alf}{\beta}$ is
given as usual by the comultiplication $\Delta(M\indf{\alf}{\beta}) =
M\indf{\alf}{\gamma}\otimes M\indf{\gamma}{\beta}$, the counit
$\epsilon(M\indf{\alf}{\beta}) = \delta\indf{\alpha}{\beta}$ and the
antipode which can be expressed using the deformed $\eps$-tensors
given explicitly in part a of the appendix:
$S(M\indf{\alf}{\beta})=\eps^{\alf\gamma}\,
M\indf{\delta}{\gamma} \, \eps_{\delta\beta}$. Together with the obvious
unimodularity condition these definitions make $M$ an $\slq$-matrix.

Since Weyl representations of the QLGr shall be constructed we impose
a unitarity condition using an antimultiplicative involution $\ast$:
$(M\indf{\alpha}{\beta})^{\ast} :=
S(M\indf{\alpha}{\beta})$, which makes $M$ an $\suq$-matrix. We call this
quantum group ${\cal A}_q$. The representations of ${\cal A}_q$ correspond
to representations built purely from the fundamental representation of
$\slq$.

To obtain the complex conjugate representations of $\slq$ an
antimultiplicative algebra morphism $k: \; {\cal A}_q \; \to \;
{\bar{\cal A}}_q \; : M \indf{\alf}{\beta} \; \to \;
{\bar M}\indc{\beta}{\alf} $ is introduced by:

\beq
k \mattwo{a}{b}{c}{d}\indf{\alpha}{\beta} \;=\;
        \mattwo{\bar a}{\bar b}{\bar c}{\bar d}\indc{\beta}{\alf}
\eeq

A comment on the indices of the complex conjugate representations has
to be made. In ordinary $SL(2,\C)$ spinor calculus one works with
dotted and undotted indices. To make contact with the classical case one
has to identify:

\beq {\bar M}\indc{\beta}{\alpha} = {\bar M}^{\dot{\alf}}_{\;\;\,\dot{\beta}}
\label{tilde-dot}
\eeq

However, in most of the calculations in this paper the calculation
with indices with tilde is more convenient since commutation
relations can more simply be formulated. If
one works with dotted indices the index-structure of the ${\hat R}$-matrices
should be altered which makes the calculus more complicated.

Sometimes the dual of the complex conjugate representation is required.
Classically this
corresponds to $M^{\dagger -1}$. Algebraically a mapping
$j: {\cal A}_q \;\to\;  {\bar{\cal A}}_q $ can be introduced \cite{PW} which
maps the fundamental to the dual complex conjugate representation by:

\beq
j\left(M\indf{\alpha}{\beta} \right) = {\bar M}\indcd{\alf}{\beta} :=
 k\left( S(M\indf{\beta}{\alf}) \right)
\eeq

It should be mentioned that $j$ can easily be extended to arbitrary
representations. Now the generating relations for ${\bar {\cal A}}_q$ can
be formulated:
\beq
{\hat R}^{-1\,\tilde{\alf}\tilde{\beta}}{}_{\tilde{\gamma}\tilde{\delta}}
{\bar M}\indcd{\gamma}{\mu}{\bar M}\indc{\delta}{\nu} \;=\;
{\bar M}\indcd{\alf}{\rho} {\bar M}\indcd{\beta}{\sigma}
{\hat R}^{-1\,\tilde{\rho}\tilde{\sigma}}{}_{\tilde{\mu}\tilde{\nu}}
\label{RTT2}
\eeq

The comultiplication $\bar{\Delta}$ and counit $\bar{\epsilon}$ on
$\bar{\cal A}_q$ are given using the mapping $k$ by $\bar{\Delta} \circ k =
(k\otimes k)\circ\Delta$ and ${}^{-} \circ \epsilon = \bar{\epsilon} \circ
k $. The antipode is again defined using the $\eps$-tensors:
$ S(\bar{M}\indc{\beta}{\alf}) = \eps_{\tilde{\alf}\tilde{\gamma}}
\bar{M}\indc{\gamma}{\delta} \eps^{\tilde{\delta}\tilde{\beta}}$.
Imposing the unimodularity and the unitarity conditions analogous to the
previous
case shows that the isomorphism holds: $\bar{\cal A}_q \simeq SU_{q^{-1}}(2)$.

As a last step commutation relations among the generators of ${\cal A}_q$ and
$\bar{\cal A}_q$ have to be fixed:

\beq
{\hat R}^{\alf\tilde{\beta}}{}_{\tilde{\gamma}\delta}
{\bar M}\indcd{\gamma}{\mu} M\indf{\delta}{\nu} \;=\;
M\indf{\alf}{\rho} {\bar M}\indcd{\beta}{\sigma}
{\hat R}^{\rho\tilde{\sigma}}{}_{\tilde{\gamma}\delta}
\label{RTT3}
\eeq

The relations (\ref{RTT1}), (\ref{RTT2}) and (\ref{RTT3}) together with
the entire Hopf structure determine the QLGr completely.

\bigskip

Important for this work are the left corepresentation spaces of the
QLGr since they are involved in the construction of spinor bases for
the $q$-deformed Poincar\'e symmetry. An irreducible right comodule
of ${\cal A}_q$ belonging to the spin $l$ representation of $\suq$ can
be defined by \cite{MAS,PW}:

\beq
 \xi_{(l)}^{\alpha} = \left[\begin{array}{c} 2l \\ l+\alf \\
            \end{array} \right]^{\half}_{q^{-2}} a^{l-\alf}c^{l+\alf}
\label{spinor}
\eeq
The definition of the $q$-binomials can be found in appendix b. The
$\xi_{(l)}^{\alf}$'s correspond to $q$-deformed symmetrized undotted
spinors of spin $l$. $\alf$ takes values in the set $\{ -l, \cdots , l \} $.
The transformation property of these undotted spinors with respect to the
QLGr are given by the comultiplication of the generators:
\beq
\Delta(\xi^{\alf}_{(l)}) = M_{(l)}{}\indf{\alf}{\beta}\otimes\xi^{\beta}_{(l)}
\label{spintrsf}
\eeq
The matrix $M_{(l)}$ is a representation of an $\suq$-matrix belonging
to spin $l$. They consist of a little $q$-Jacobi polynomial and certain
powers of the quantum group generators. Details can be found in \cite{MAS}.
Applying the mappings $k$ and $j$ introduced above one obtains the other
corepresentations:
\beq
\bar{\xi}_{(l)\,{\tilde{\alf}}}:= k(\xi^{\alf}_{(l)}), \quad
\bar{\xi}_{(l)}^{\tilde{\alf}}:= j(\xi^{\alf}_{(l)}), \quad
     \xi_{(l)\,\alf} = k\circ j (\xi_{(l)}^{\alf}).
\label{spinors}
\eeq
The four corepresentation spaces (\ref{spinor}) and (\ref{spinors}) play
the same roles as their undeformed analogues do in the van der Waerden
spinor calculus. Their transformation properties are obvious from
(\ref{spintrsf}).

A ge\-ne\-ral co\-re\-pre\-sen\-ta\-tion of the QLGr is then given by the
(non-commuting)
product: $\bar{\xi}_{(l_1)}\xi_{(l_2)}$. This is equivalent to the
Weyl representation of the undeformed Lorentz group with highest
weights $(l_1,l_2)$.

\bigskip

The last point in this section is the Iwasawa decomposition of the
QLGr \cite{PoWo}. In the undeformed case a Lorentz transformation can
be decomposed into a product of a pure space rotation and a pure boost.
This is valid also in the quantum case. It holds in the fundamental
representation that $ M= w_R \, w_B $, with $(w_R)^i_j$ being a $\suq$-matrix.

\beq
(w_B)\indf{\alf}{\beta} := \mattwo{\rho}{z}{0}{\rho^{-1}}\indf{\alf}{\beta}
\label{boostdef}
\eeq
is a $\slq$-matrix and its generating relations can be obtained by inserting
the
entries into (\ref{RTT1}), (\ref{RTT2}) and (\ref{RTT3}). This
decomposition can be extended to an arbitrary representation. The Hopf
structure
is the same as above for an $\slq$ matrix. It should be
mentioned that because of the unitarity of $w_R$ it holds that
$k(M) = k(w_B) S(w_R)$. Since the generating relations of $w_B$ will
become important in this work they are listed here.
\beq
\rho\rho^{\ast} \;=\; \rho^{\ast}\rho, \quad \rho z \;=\; q z \rho, \quad
z \rho^{\ast} \;=\; q^{-1} \rho^{\ast} z,  \quad
zz^{\ast} \;=\; z^{\ast} z-q\lam
    \left( (\rho^{\ast}\rho)^{-1} -\rho^{\ast}\rho \right).
\label{boostrel}
\eeq

This completes the study of the QLGr for the purposes of this work. More
details about the $q$-deformed Lorentz symmetry can be found in
\cite{PoWo,PW,TAK}.

\section{$q$-deformed one-particle states}

\bigskip

In \cite{PSW,TALK} it has been shown that unitary Irreps of the
$q$-deformed Poincar\'e symmetry \cite{POI} can be constructed for
the massive and the massless case. We will state briefly the results
here mainly for the first case.

The $q$-deformation of the Poincar\'e algebra was obtained adding an
inhomogeneous part to the $q$-deformed Lorentz algebra \cite{LOR} which
consists merely of the $({1\over 2},{1\over 2})$-
corepresentation of the QLGr, i.e. a bispinor ${\bar{\xi}_{(1/2)}^{\dot{\alf}}
\xi_{(1/2)}^{\beta}}$. The vector components are labeled
$\left(A,B,C,D\right)$. Using the results of the previous section one
obtains the commutation relations of these components:

\beq
\beqcol
A B &=& B A - q^{-1} \lambda C D + q \lambda D^2, \\
A C &=& C A + q \lambda A D, \\
A D &=& q^{-2} D A,
\eeqcol \qquad
\beqcol
B C &=& C B - q^{-1} \lambda B D, \\
B D &=& q^2 D B, \\
C D &=& D C.
\eeqcol
\label{ppalg}
\eeq
The components behave under complex conjugation $k$ which is just denoted
by a bar:
\beq
{\overline A}=B,\qquad {\overline B}=A,\qquad {\overline C}=C, \qquad
     {\overline D}=D.
\eeq

A $q$-analogue of a metric tensor exists which shows Minkowskian signature:

\beq
g^{IJ} = {\left( \begin{array}{cccc}
            0  &  1  &  0  &  0  \\
         q^{-2}&  0  &  0  &  0  \\
            0  &  0  &-q\lam & -1 \\
            0  &  0  &  -1  & 0  \\   \end{array} \right)}^{IJ},
\label{metric}
\eeq
with inverse $g_{IJ} g^{JK} = \delta_I^K$. The length of the momentum
four vector
\beq
P^2 = -(q^2+1)^{-1}P^I g_{IJ}P^J = q^{-2} CD -AB =: M^2
\label{mass}
\eeq
is a Casimir in the $q$-deformed Poincar\'e algebra and can be interpreted
as mass. The convention $P^A := A$ is chosen. In \cite{PSW} it turned out
that a unitary Irrep of the algebra is
classified by the eigenvalue of $M^2$ and the states are labeled by the
real eigenvalues of the energy- and $z$-component of the $q$-four vector:
$P^0=q(q+q^{-1})^{-1} (C+D)$ and $P^z = (q+q^{-1})^{-1}
(qD-q^{-1}C)$ resp., the third component $l$ of the orbital angular
momentum operator $T^3$ and an additional parameter $r$ which takes values
$0$ or $1$. It should be mentioned that one can not take a basis in which
all components of the four vector are diagonal as can be seen from the algebra
(\ref{ppalg}). A general Hilbert space state
$ \ket{n,N,l,r,F}=:\ket{{\cal P}}$ is labeled by the integer eigenvalues of
the diagonal generators:
\beq
\beqcol
m^2 &=& d_0^2 q^{2F}, \\
t_3 &=& q^{-1}[2l]_{q^{-2}}, \\
\eeqcol \qquad
\beqcol
p^0 &=& d_0 {{q^{1-r}}\over {q+q^{-1}}}\left( q^{2(N+1)} +
                       q^{2(F-N+r)} \right), \\
p^z &=& d_0{{q^{-r}}\over {q+q^{-1}}}
        \left(q^{2n} - {{q^{2(N+1)} + q^{2(F-N+r)}}\over{q^2+1}} \right).
\eeqcol
\label{eigenv}
\eeq
$d_0$ is a real universal parameter whose sign characterizes the
sign of the energy as in the undeformed representation theory.

An important result of this analysis is that the mass is quantized
in terms of the deformation parameter, the quantum number $F$ classifies
the different Irreps. (\ref{eigenv}) shows that the spectra of energy
and momentum are discretized in the deformed case. The norm of the states
is obvious: $\scpro{{\cal P}^{\prime}}{{\cal P}} =
\delta_{{\cal P}^{\prime}{\cal P}}$. In the sense of Wigner \cite{WIG} these
unitary Irreps of the $q$-deformed Poincar\'e algebra can be considered
as $q$-deformed massive relativistic one-particle states.

The construction of \cite{PSW} did not allow the incorporation of spin,
although
the analysis showed that the stability subgroup inducing the
massive representations is $\suq$. Therefore in \cite{PW} covariant-
or spinor- representations have been constructed. A general spinor state
is given by:
\beq
\ket{{\cal P};\, l_1,\alf; \, l_2, \beta } \;: =\;
   \ket{{\cal P}} \otimes\bar{\xi}^{\tde{\alf}}_{(l_1)}\xi^{\beta}_{(l_2)}
\label{spinorbas}
\eeq
These representations are then induced by a representation $(l_1,l_2)$ of
the QLGr. For example a pure undotted spinor belonging to spin $l_2$ is
just $\ket{{\cal P}; 0,0; l_2,\beta} $. These spinor states are
of course not orthonormal by themselves as it has been shown in section 2.
This problem is addressed in the next section.

Using different techniques massless representations can be obtained
\cite{TALK}. In
this case an additional quantum number, the $q$-analogue of the helicity,
occurs to label the unitary Irreps. From the eigenvalue of the mass
in (18) one sees that in this case $F \to - \infty$ keeping $N$ fixed.
\section{$q$-deformed Dirac equation}

\bigskip

As has been argued in section 2 the wave operator can be extracted
merely from the inner product of two spinor states. Hence there is
a need for a inner product for the $q$-deformed spinor states introduced
in (\ref{spinorbas}).

In general a scalar product $\scpro{\cdot}{\cdot}$ for a star-Hopf algebra
$(H,\Delta,\epsilon, S,\ast)$ and a finite dimensional left-H-comodule $V$ is
given by the mapping \cite{MAS,PW}:
\beq
( \cdot, \cdot )_L \; ; \; (H\otimes V)\times (H\otimes V) \;\to\; H \; :\;
(a\otimes \xi )\times (b\otimes \eta ) \mapsto ab^{\ast}
\scpro{\xi}{\eta}_L
\label{scprodef}
\eeq
As long as the quantum group is compact and unitarily represented the
inner product is given by the Haar measure and bi-invariant. The following
normalization is introduced:
\beq
\scpro{\xi^{\alf}_{(l_1)}}{\xi^{\beta}_{(l_2)}}_L =\delta_{l_1 l_2}
      \delta^{\alf\beta}, \qquad
\scpro{\bar{\xi}^{\dot{\alf}}_{(l_1)}}{\bar{\xi}^{\dot{\beta}}_{(l_2)}}_L
   = \delta_{l_1 l_2}\delta^{\dot{\alf}\dot{\beta}} .
\label{spinnorm}
\eeq
Since the QLGr is not compact it can not be expected that the so
constructed inner product is still invariant. This means that when
transforming
the expressions in (\ref{spinnorm}) with a $q$-Lorentz transformation
(\ref{spintrsf}) in the Iwasawa decomposition of section 3 it can be seen from
(\ref{scprodef}) that a term coming from the pure boost part $w_B$
survives. The pure rotation part $w_R$ vanishes since the $\xi$'s and
$\bar{\xi}$'s are unitary corepresentations of ${\cal A}_q$ and
$\bar{\cal A}_q$
respectively. It will be shown that this surviving part gives the
wave operator in the sense of section 2. It corresponds to the
expression ${\cal D}\left( { {\sigma . p} \over {m}}\right)$
occuring in (\ref{ipr2}).

Before it is shown how the procedure works in the general case a
few special physically interesting cases are treated
in order to make the techniques transparent.

The first example is the case of spin $1/2$. This
will lead to the $q$-deformed Dirac operator. The corresponding
representations of the $q$-deformed Poncar\'e symmetry are in this
case induced by the representation $\left( ({1\over 2},0 ) \oplus
(0, {1\over 2}) \right)$ of the QLGr. For the construction of the
wave operators it is sufficient to consider only the pure Lorentz inner
products. The coaction is applied to the spinors in (\ref{spinnorm}):
\begin{eqnarray}
\scpro{\Delta(\xihf^{\alpha})}{\Delta(\xihf^{\beta})}_L &=&
(w_B)\indf{\alf}{\gamma} \left( ( w_B)\indf{\beta}{\rho} \right)^{\ast}
\scpro{\xihf^{\gamma}}{\xihf^{\rho}}_L \\
\: \: {} &=&
(w_B)\indf{\alf}{\gamma} \left( (w_B)\indf{\beta}{\gamma} \right)^{\ast}
=:\Pxi{\half}\indf{\alf}{\tilde{\beta}}
\label{Dirxi}
\end{eqnarray}
An analogous procedure for the dual complex conjugate representation
can be performed yielding:
\beq
\scpro{\Delta(\bar{\xi}_{(\half )}^{\tilde{\alf}})}
{\Delta({\bar{\xi}_{(\half )}}^{\tilde{\beta}})}
=: \Pxib{\half}\indf{\tilde{\alf}}{\beta}
\label{Dirxib}
\eeq

The matrices occuring in (\ref{Dirxi}) and (\ref{Dirxib}) read
explicitly in terms of the generators of the boost matrices $w_B$:
\beq
\Pxi{\half}\indf{\alf}{\tilde{\beta}} =
 \mattwo{\rho\rho^{\ast}+z^{\ast}z}{z\rho^{-1 \ast}}{\rho^{-1}z^{\ast}}
        {\rho^{-1}\rho^{-1\ast}} \indf{\alf}{\tilde{\beta}} , \quad
\Pxib{\half}\indf{\tilde{\alf}}{\beta} =
 \mattwo{\rho^{-1\ast}\rho^{-1}}{-q^{-1}\rho^{-1\ast}z}
        {-q^{-1}z^{\ast}\rho^{-1}}{q^{-2}z^{\ast}z+\rho^{\ast}\rho}
        \indf{\tilde{\alf}}{\beta}.
\label{Direxp}
\eeq

It is interesting that the entries of $P_{\xi_{1/2}}$ and
$P_{\bar{\xi}_{1/2}}$ can be algebraically identified with the generators
of the momentum part of
the $q$-deformed Poincar\'e algebra (\ref{ppalg}) normalized by the
mass:.
\beq
\beqcol
M^{-1} A &=& - z^{\ast}\rho^{-1}, \\
M^{-1} B &=& - \rho^{-1\ast} z ,
\eeqcol \qquad \qquad
\beqcol
M^{-1} C &=& q^2 \rho^{\ast}\rho +z^{\ast}z ,  \\
M^{-1} D &=& \rho^{-1\ast} \rho^{-1} .
\eeqcol
\label{identi}
\eeq
For later use it is also mentioned that
$M^{-1}(C-q\lam D) = \rho\rho^{\ast}+zz^{\ast}$. It is easy to see that the so
constructed quantities obey the algebra (\ref{ppalg}) of the $q$-deformed
Minkowski four vector using the relations (\ref{boostrel}). One can now
rewrite (\ref{Direxp}) :
\beq
\Pxi{\half}\indf{\alf}{\tde{\beta}} = M^{-1}\;
 \mattwo{C-q\lam D} {-q B}
        {-q A}{D}      \indf{\alf}{\tde{\beta}}, \qquad
\Pxib{\half}\indf{\tde{\alf}}{\beta} =  M^{-1}
 \mattwo{D}{q^{-1} B}
        {q^{-1} A}{q^{-2} C} \indf{\tde{\alf}}{\beta}.
\label{diracop}
\eeq
These wave operators coincide with the $q$-deformed Dirac operators
constructed in \cite{ARNE} where an approach with a $q$-deformed
Clifford algebra has been used. The correspondence is:
\beq
P_{\xi_{(\half )}} \sim {{\sigma .p}\over {m}}, \qquad
P_{\bar{\xi}_{(\half )}} \sim {{\bar{\sigma}.p}\over {m}}.
\eeq

The $\sigma$'s denote the $q$-deformed Pauli matrices in the basis of
\cite{ARNE}.

The $q$-deformed Dirac equation on the wave functions $\psi^{\beta}({\cal P})$
and $\bar{\phi}^{\tde{\alf}}({\cal P})$ then is:
\beq
\beqcol
 \Pxib{1/2}\indf{\tde{\alf}}{\beta} \psi^{\beta}({\cal P}) &=&
 \bar{\phi}^{\tde{\alf}}({\cal P}),   \\
 \Pxi{1/2}\indf{\beta}{\tde{\alf}} \bar{\phi}^{\tde{\alf}}({\cal P}) &=&
  \psi^{\beta}({\cal P}),
\eeqcol  \qquad\qquad
\beqcol
 P^2 \psi^{\beta}({\cal P}) &=& m^2 \psi^{\beta}({\cal P}), \\
 P^2 \bar{\phi}^{\tde{\alf}}({\cal P}) &=&
		m^2   \bar{\phi}^{\tde{\alf}}({\cal P}).
\eeqcol
\label{diraceq}
\eeq
The mass shell conditions are obtained by iterating the $q$-deformed
Dirac operators and are of course a consequence of the representation
theory of the $q$-deformed Poincar\'e symmetry. For further properties
of the $q$-deformed Dirac operators the reader is referred to \cite{ARNE}.

\section{$q$-deformed spin 1 Joos-Weinberg- and Maxwell- equations}

\bigskip

To construct a wave equation for a spin one representation one has
the choice between inducing with a representation $\left( (1,0) \oplus
(0,1) \right)$ or $(\half , \half )$ of the QLGr. This section deals with
the first case which provides the simplest non-trivial example of a
$q$-deformed wave equation in the Joos-Weinberg basis \cite{JOOS,WEINB}. The
second case will lead to the $q$-deformed Proca equation and is addressed in
the next section.

\bigskip
\goodbreak

{\bf 6.1 Spin 1 Joos-Weinberg equation}

\bigskip

We use the comodules $\xi_{(1)}^a$ and $\bar{\xi}_{(1)}^{\tde{b}}$ of the
QLGr. By construction these comodules have 3 degrees of freedom each and
the indices run through the set $\{-1,0,+1 \}$. These representations
correspond to the product of two fundamental representations which have
been symmetrized and from which the trace part has been eliminated.

{}From the inner product
(\ref{spinnorm}), the spin 1 representation of the $q$-deformed
boost matrices $w_B$ (see appendix c) and the identifications
(\ref{identi}) of the preceding section the following can be deduced:
\begin{eqnarray}
\scpro{\Delta(\xi_{(1)}^a)}{\Delta(\xi_{(1)}^b)}_L &=& (w_B)\indf{a}{c}
\left( (w_B)\indf{b}{c} \right)^{\ast}=: \Pxi{1}\indf{a}{\tde{b}} =  \\
&=& M^{-2}\mattre{ DD }{ -q\omega DA }{ q^3 AA}
{ -q\omega BD }{ q^2\omega BA + M^2}{ -q\omega(\cld)A }
 { q^3 BB }{ -q^2\omega B(\cld) }{ (\cld)^2-q^3\lambda BA  }\indf{a}{\tde{b}}
\nonumber
\label{JWxi}
\end{eqnarray}

The abbreviation $\omega=\sqrt{1+q^{-2}}$ has been used. From the analysis
of $\bar{\xi}_{(1)}^{\tde{a}}$ one obtains:
\beq
\Pxib{1}\indf{\tde{b}}{a}= M^{-2}
\mattre{q^{-4}CC+q^{-3}\lam AB}{q^{-3}\omega AC}{q^{-3} AA}
  {q^{-3}\omega CB}{\omega^2 AB+M^2}{\omega AD}
  {q^{-3}BB}{\omega DB}{DD}    \indf{\tde{b}}{a}
\label{JWxib}
\eeq
Using these wave operators the spin 1 wave equation can be formulated:
\beq
\beqcol
\Pxi{1}\indf{a}{\tde{b}} \bar{\chi}^{\tde b}({\cal P}) &=& \psi^a ({\cal P}),\\
\Pxib{1}\indf{\tde{b}}{a} \psi^a({\cal P}) &=& \bar{\chi}^{\tde b}({\cal P}),
\eeqcol \qquad\qquad
\beqcol
P^2 \bar{\chi}^{\tde b}({\cal P}) &=& m^2 \bar{\chi}^{\tde b}({\cal P}), \\
P^2 \psi^a({\cal P}) &=& m^2 \psi^a({\cal P}) .
\eeqcol
\label{spin1eq}
\eeq
It will now be shown that the system of spin 1 wave equations
(\ref{spin1eq}) can equivalently be written using the $q$-deformed
Dirac operators (\ref{diracop}). One can reexpress the fields $\bar{\chi}$
and $\psi$ as follows:
\begin{eqnarray}
\bar{\chi}^{\tde 1} = \bar{\varphi}^{\tde{2}\tde{2}}, \qquad
&\bar{\chi}^{\tde 0} = (q\omega)^{-1}\left(\bar{\varphi}^{\tde{1}\tde{2}} +
                       q^{-1}\bar{\varphi}^{\tde{2}\tde{1}} \right),\qquad
&\bar{\chi}^{-\tde{1}} = \bar{\varphi}^{\tde{1}\tde{1}},                \\
\psi^1 = \phi^{22}, \qquad
& \psi^0 = \omega^{-1}\left( \phi^{12}+q^{-1}\phi^{21} \right), \qquad
&\psi^{-1} = \phi^{11}.
\label{spinid}
\end{eqnarray}
It is possible to show the following equivalence by direct inspection:
\beq
\Pxi{1}\indf{a}{\tde b} \simeq W^{\alf\rho}{}_{\tde{\sigma}\tde{\delta}}
:= q {\hat R}^{-1\;\tde{\beta}\rho}{}_{\gamma\tde{\sigma}}
 \Pxi{1/2}\indf{\alf}{\tde{\beta}} \Pxi{1/2}\indf{\gamma}{\tde{\delta}}
\label{waveopid}
\eeq
It is important to note that the matrix ${\hat R}^{-1}$ has the purpose of
commuting the index ${\tde{\beta}}$ of the first Dirac operator with the
index $\gamma$ of the second one in order to have the index structure
as indicated in the object $W$. This index permutation occurs every time
when one expresses a $q$-deformed wave operator in terms of
more than one $q$-deformed Dirac operator. One then obtains a $q$-deformed
wave equation in 2-spinor indices:
\beq
 W^{\alf\rho}{}_{\tde{\sigma}\tde{\delta}} \, \bar{\varphi}^{\tde{\delta}
\tde{\sigma}}({\cal P}) = \phi^{\alf\rho}({\cal P})
\label{JWeq}
\eeq
The same could have been done with the second pair of equations in
(\ref{spin1eq}).

\bigskip
\bigskip

{\bf 6.2 $q$-deformed Maxwell equations}

\bigskip

It will now be shown that when setting the mass equal to zero in the
wave equations (\ref{spin1eq}) these equations can be rewritten as
first order equations. These first order equations turn out to be
$q$-analogues of the Maxwell equations formulated in terms of an
electromagnetic spinor cf. \cite{PR}. Performing this reduction
and using (\ref{spinid}) one obtains:
\beq
\beqcol
(\cld) \:\bar{\varphi}^{\tde{1}\tde{1}} -qB \:\bar{\varphi}^{\tde{2}\tde{1}}
&=& 0,\\
\qquad
(\cld) \:\bar{\varphi}^{\tde{1}\tde{2}} -qB \:\bar{\varphi}^{\tde{2}\tde{2}}
&=& 0,
\eeqcol \qquad
\beqcol
-qA \:\bar{\varphi}^{\tde{1}\tde{1}} + D \:\bar{\varphi}^{\tde{2}\tde{1}} &=&
0, \\
-qA \:\bar{\varphi}^{\tde{1}\tde{2}} + D \:\bar{\varphi}^{\tde{2}\tde{2}} &=&
0.
\eeqcol
\label{MW1}
\eeq
This system can be rewritten in terms of a single $q$-deformed Dirac operator:
\beq
\Pxi{1/2}\indf{\alf}{\tde{\beta}} \bar{\varphi}^{\tde{\beta}\tde{\gamma}}
({\cal P}_{m^2=0})
        = 0
\label{MW2}
\eeq
Again the same can be done for the field $\phi^{\alf\beta}$. When an
appropriate reality condition is stated, e.g. $\bar{\varphi}^{\tde{\alf}
\tde{\beta}} = j(\phi^{\alf\beta}) $ then (\ref{MW1}) or
(\ref{MW2}) can be considered as the $q$-deformed Maxwell equations in
momentum space. It is surely possible to change the basis in such a way
that the equations can be written in terms of electric- and magnetic-
fields.
\section{$q$-deformed Proca equation}

\bigskip

A more familiar description of a spin 1 field is that of an Irrep of
the Poincar\'e group which is induced by the $(\half ,\half )$ representation
of the Lorentz group. The wave equation belonging to this case is
the Proca equation \cite{PRO}. It will be shown that a
$q$-deformation of
this equation is possible. However, one encounters some technical
problems in the deformed case.

Inducing with the $(\half ,\half )$ representation of the QLGr means to
work with a $q$-deformed four vector $\bar{\xi}_{(1/2)}^{\tde{\alf}}
\xi_{(1/2)}^{\beta}$. In the following the representation labeling of
$\xi$ and $\bar{\xi}$ will be omitted. Inserting the transformed four vector
into the inner product (\ref{scprodef}) yields:
\beq
\Pi := \scpro{\Delta(\bar{\xi}^{\tde{\alf}})\Delta(\xi^{\beta})}
             {\Delta(\bar{\xi}^{\tde{\gamma}})\Delta(\xi^{\delta})}_L
= (\bar{w}_B)\indcd{\alf}{\alf^{\prime}} (w_B)\indf{\beta}{\beta^{\prime}}
 (\bar{w}_B)\indc{\delta^{\prime}}{\delta} (w_B)\indfd{\gamma^{\prime}}{\gamma}
 \langle \bar{\xi}^{\alf^{\prime}} \xi^{\beta^{\prime}}
 \bar{\xi}_{\tde{\delta}^{\prime}} \xi_{\gamma^{\prime}}\rangle
\label{pr1}
\eeq
The goal will be to permute the first matrix in (\ref{pr1}) to the immediate
left
of the matrix $(w_B)\indfd{\gamma^{\prime}}{\gamma}$ because then
the inner product can be evaluated and the remaining expression can be
written in terms of $q$-deformed Dirac operators which is more transparent.
Using the ordinary ${\hat R}$-matrix calculus one gets:
\beq
\Pi = {\hat R}^{-1\:\tde{\alf}{\beta}}{}_{\rho \tde{\alf^{\prime}}}
      {\hat R}^{-1\:\tde{\delta}^{\prime}\tde{\alf}^{\prime}}{}_{\tde{\sigma}
               \tde{\delta}}
(w_B)\indf{\rho}{\beta^{\prime}} (\bar{w}_B)\indc{{\delta}{\prime\prime}}
                                         {{\delta}^{\prime}}
(\bar{w}_B)\indcd{\sigma}{\nu^{\prime}}(w_B)\indfd{\gamma^{\prime}}{\gamma}
\langle\xi^{\beta^{\prime}}\bar{\xi}_{\tde{\delta}^{\prime\prime}}
\bar{\xi}^{\tde{\nu}^{\prime}} \xi_{\gamma^{\prime}} \rangle.
\label{pr2}
\eeq
The commutation of the objects remaining in the inner product has
not been written out explicitly only the result of the $q$-permutation is
indicated in the bracket. Now the expression in the inner
product can be evaluated:
\beq
\langle\xi^{\beta^{\prime}}\bar{\xi}_{\tde{\delta}^{\prime\prime}}
\bar{\xi}^{\tde{a}^{\prime}} \xi_{\gamma^{\prime}} \rangle
=\scpro{\xi^{\beta^{\prime}}}{\xi^{\delta^{\prime\prime}}}_L
\scpro{\bar{\xi}^{\tde{a}^{\prime}}}{\bar{\xi}^{\tde{\gamma}^{\prime}}}_L
=\delta^{\beta^{\prime}}_{\tde{\delta}^{\prime\prime}}
   \delta^{\tde{a}^{\prime}}_{\gamma^{\prime}}
\label{pr3}
\eeq
The last term in this equation is meant only symbolically. Now
$\Pi$ can be expressed in terms of $q$-deformed Dirac operators:
\beq
\Pi= {\hat R}^{-1\:\tde{\alf}{\beta}}{}_{\beta^{\prime} \tde{\alf^{\prime}}}
   {\hat R}^{-1\:\tde{\delta}^{\prime}\tde{\alf}^{\prime}}{}_{\tde{\rho}
		  \tde{\delta}}
 \Pxi{1/2}\indf{\beta^{\prime}}{\tde{\delta}^{\prime}}
 \Pxib{1/2}\indf{\tde{\rho}}{\mu}
\label{pr4}
\eeq
Using the $\eps$-tensors of the appendix the indices in the previous
expression for $\Pi$ can be raised and lowered in such a way that they can be
reexpressed in a four vector index:
\beq
\Pi^{IJ} = -P\indfd{\beta}{\tde{\alf}}P\indfd{\delta}{\tde{\gamma}} \, - \,
            g\indfd{\beta}{\tde{\alf}}{}\indfd{\delta}{\tde{\gamma}}
\; = \; - P^I\,P^J \: - \: g^{IJ},
\label{pr5}
\eeq
with
\beq
P\indfd{\beta}{\tde{\alf}} = \mattwo{C}{-A}{-B}{D}\indfd{\beta}{\tde{\alf}}
=: P^I,
\label{pdef}
\eeq
and the tensor $g^{IJ}$ is the $q$-deformed metric (\ref{metric}). The capital
indices run through the set ${A,B,C,D}$.
However, the object $\Pi^{IJ}$ is not yet the Proca operator. This is a
result of the fact that when the Dirac operators are subsequently applied to a
four-vector field $A^{\dot{\alf}{\beta}}$ after the
action of the first operator a symmetrization in the dotted indices has to
be performed in
order to avoid the appearance of a scalar. This could have already been
implemented in the above calculation but it is now explained in a
different way.

One uses the symmetrizer which comes out of the characteristic
polynomial for ${\hat R}$. It is convenient to express it as:
$ S = 1 + (q+q^{-1})^{-1} \eps\eps $. The term with the $1$ gives just
$\Pi^{IJ}$ while the second has to be applied to the first $q$-Dirac
operator and the vector field to which it is applied. Then the
second Dirac operator is multiplied from the left and the spinor indices are
brought in an order comparable to that in (\ref{pr5}). This gives
an additional term $q(q+q^{-1})^{-1}P^I\,P^J$ which has to be added to
$\Pi^{IJ}$.

Then the wave operator applied to a $q$-four vector field
$A_J({\cal P})$ gives:
\beq
\left( - {1\over{(q^2+1)M^2}} P^I\, P^J \: - \: g^{IJ} \right)\; A_J({\cal P})
\; = \; A^I({\cal P})
\label{procaeq1}
\eeq
The operator on the left hand side of the previous equation will be
abbreviated by $Pr^{IJ}$. Multiplying equation (\ref{procaeq1}) from the left
with $P^K \, g_{KJ}$ and using
(\ref{mass}) gives the $q$-analogue of the Proca equation:
\beq
 P^I\: g_{IJ} \: A^{J}({\cal P}) \; = \; 0
\label{procaeq2}
\eeq
This equation simply means that the scalar component with respect
to $\suq$ of the $q$-deformed four vector has to vanish.

\section{$q$-deformed Rarita-Schwinger equation}

\bigskip

A physical field of spin $3\over 2$ is suitably described by a representation
of the Poincar\'e group which is induced either by a $\left( ({3\over 2},0)
\oplus (0,{3\over 2}) \right)$- or a $\left( (1,\half ) \oplus (\half , 1)
\right)$- representation of the Lorentz group. The first possibility
leads to a further example of a Joos-Weinberg equation whose $q$-deformation
will be treated in the next section in a general setting. The second
choice leads to the Rarita-Schwinger equations \cite{RS}.

The starting point for the $q$-deformation of the Rarita-Schwinger equations
is the product of three fundamental representations of $\slq$:
$\bar{\xi}^{\tde{\alf}} \bar{\xi}^{\tde{\beta}} \xi^{\gamma}$. In this
product an additional $(0,\half )$ representation of the QLGr occurs which
can be eliminated by the condition $\eps_{\tde{\alf}\tde{\beta}}
\bar{\xi}^{\tde{\alf}} \bar{\xi}^{\tde{\beta}} \xi^{\gamma} = 0 $. Now
the procedure of section 7 can be repeated in this more complicated case.
This leads to a threefold product of $q$-deformed Dirac operators where
the spinor indices of the operators have to be arranged in a naturally
appropriate way. The Rarita-Schwinger operator ${\cal R}_1$ which occurs
can be written symbolically as:
\beq
{\cal R}_1 = r_{12}r_{23}r_{12}r_{34}r_{23}r_{45}
(P_{\xi})\indf{\bf{1}}{\tde{\bf{2}}} (P_{\bar{\xi}})\indf{\tde{\bf{3}}}
{\bf{4}} (P_{\bar{\xi}})\indf{\tde{\bf{5}}}{\bf{6}}
\label{rs1}
\eeq
This equation requires some explanation. The
$r_{ij}$'s are permutation operators between the $i$th and $j$th
place in the sixfold product of spinor indices coming from the Dirac
operators. They consist in principle of ${\hat R}$ or ${\hat R}^{-1}$
matrices. The boldface suffixes on the Dirac operators denote just the
type of $\slq$ index entering. The explict expression of (\ref{rs1}) is
not very transparent. It can be found in appendix d.

It can be shown that ${\cal R}_1$ admits a decomposition into a
$q$-deformed Proca- and a Dirac- operator:
\beq
{\cal R}_1 = r_{12} r_{23} r_{34} r_{45} (Pr)\indf{\tde{\bf 1}}{\bf{2}}
             {}\indf{\tde{\bf 3}}{\bf{4}}
           (P_{\bar{\xi}})\indf{\tde{\bf{5}}}{\bf{6}}
\label{rs2}
\eeq
A similar procedure can be performed for the $(\half , 1)$ product
$\bar{\xi}^{\tde{\alf}} \xi^{\beta} \xi^{\gamma}$. One eliminates
the unwanted $( \half , 0)$ component as above by :
$\eps_{\beta\gamma}\bar{\xi}^{\tde{\alf}}\xi^{\beta}\xi^{\gamma} =0$. This
leads to a second part of the Rarita-Schwinger operator:
\beq
{\cal R}_2 = r_{23} r_{12} r_{34} r_{23} (P_{\xi})\indf{\bf{1}}{\tde{\bf {2}}}
             (Pr)\indf{\tde{\bf{3}}}{\bf{4}} {}\indf{\tde{\bf{5}}}{\bf{6}}
\label{rs3}
\eeq
Thus one gets two equations in the parity components
$\chi^{\tde{\alf}\tde{\beta}\gamma}({\cal P})$ and
$\psi^{\tde{\rho}\sigma\lambda}({\cal P})$ of the Rarita-Schwinger field:
\begin{eqnarray}
{\cal R}_1 {}^{\tde{\alf}\tde{\beta}\gamma}{}_{\tde{\rho}\sigma\lambda}
\psi^{\tde{\rho}\sigma\lambda}({\cal P}) &=&
\chi^{\tde{\alf}\tde{\beta}\gamma}({\cal P})  \\
{\cal R}_2 {}^{\tde{\rho}\sigma\lambda} {}_{\tde{\alf}\tde{\beta}\gamma}
\chi^{\tde{\alf}\tde{\beta}\gamma}({\cal P}) &=&
 \psi^{\tde{\rho}\sigma\lambda}({\cal P})
\label{rs4}
\end{eqnarray}
The mass shell conditions for the fields are obvious. These equations can be
rewritten in terms of linear equations.
The way to obtain these is somewhat technical. The key point is
to reorder the spinor indices of ${\cal R}_1$, ${\cal R}_2$ and the
fields and to multiply the wave operators appropriately from the
left by $q$-deformed Dirac- and Proca- operators. As in the case of the
vector field in the previous section a dotted and an undotted index on
the fields are combined to a four vector index. One obtains:
\beq
\beqcol
P^I g_{IJ} \chi^{J\,\tde{\alf}}({\cal P})  & = & 0,  \\
(P_{{\xi}_{(1/2)}})\indf{\beta}{\tde{\alf}}  \chi^{\tde{\alf}\, J}({\cal P})
&=&          \psi^{\beta\, J}({\cal P})
\eeqcol \qquad
\beqcol
P^I g_{IJ} \psi^{J \, \beta}({\cal P}) & = &  0 ,     \\
(P_{{\bar{\xi}}_{(1/2)}})\indf{\tde{\alf}}{\beta} \psi^{\beta\, J}({\cal P})
&=&              \chi^{\tde{\alf}\, J}({\cal P}).
\eeqcol
\label{rseq}
\eeq
Again the result is that the $q$-deformed wave equations look
structurally like the undeformed ones besides the complicated index
rearrangements using ${\hat R}$-matrices. Nevertheless the key point
is that the entries of the wave operators are generators of a
non-commutative algebra.

\section{The general case}

\bigskip

So far $q$-deformed wave equations have been constructed for low
dimensional but physically important inducing representations of the
QLGr. It is possible to generalize the constructions to an arbitrary
inducing representation. This is most easily done for the Joos-Weinberg
equations, i.e. for an inducing representation $\left( (j,0) \oplus (0,j)
\right)$ which is a direct generalization of sections 5 and 6.1.

One considers the corepresentations $\xi_{(j)}^{\alf}$ and
$\bar{\xi}_{(j)}^{\tde{\beta}}$ introduced in (\ref{spinor}) and
(\ref{spinors}) and studies the inner product of the transformed
spinors:
\begin{eqnarray}
\Pxi{j}\indf{\alf}{\tde{\beta}} &:=&
          \scpro{\Delta(\xi_{(j)}^{\alf})}{\Delta(\xi_{(j)}^{\beta})}_L =
       (w_B^{(j)})\indf{\alf}{\gamma} \left( (w_B^{(j)})\indf{\beta}{\gamma}
        \right)^{\ast}       \\
\Pxib{j}\indf{\tde{\alf}}{\beta} &:= &
       \scpro{\Delta(\bar{\xi}_{(j)}^{\tde{\alf}})}
             {\Delta(\bar{\xi}_{(j)}^{\tde{\beta}})}_L =
   (\bar{w}_B^{(j)})\indcd{\alf}{\gamma}
\left( (w_B^{(j)})\indcd{\beta}{\gamma} \right)^{\ast}
\label{gen1}
\end{eqnarray}
The higher dimensional representations of the boost matrices $w_B$ can
be taken from part c of the appendix and the entries of the $q$-deformed
wave operators can be identified with the generators of the
non-commu\-ta\-tive Minkowski space using (\ref{identi}).
Taking fields $\psi^{\beta}({\cal P})$ and $ \chi^{\tde{\alf}}({\cal P})$,
the indices $\alf$ and $\beta$ taking values in the set
$\{ -j,\cdots , j \}$, which are Irreps of the $q$-deformed Poincar\'e
symmetry and therefore are already on shell one can formulate the general
$q$-deformed
Joos-Weinberg wave equations:
\beq
\beqcol
\Pxi{j}\indf{\alf}{\tde{\beta}} \bar{\chi}^{\tde \beta}({\cal P}) &=&
 \psi^{\alf} ({\cal P}), \\
\Pxib{j}\indf{\tde{\beta}}{\alf} \psi^{\alf}({\cal P}) &=&
 \bar{\chi}^{\tde \beta}({\cal P}),
\eeqcol \qquad\qquad
\beqcol
P^2 \bar{\chi}^{\tde \beta}({\cal P}) &=& m^2 \bar{\chi}^{\tde \beta}({\cal
P}), \\
P^2 \psi^{\alf}({\cal P}) &=& m^2 \psi^{\alf}({\cal P}) .
\eeqcol
\label{geneq}
\eeq
It can moreover be proven by induction from the results of
sections 6, 7, 8 and especially equation (\ref{waveopid}) that one can start
equivalently with a
tensor product of fundamental representations which have to be made
irreducible by symmetrizing and removing the traces. This leads to a
product of $q$-deformed
Dirac operators in which the indices have to be arranged in a correct
way. An example is shown in part d of the appendix. These expressions
are not transparent and therefore they are not given explicitly for
the higher spins. The $q$-deformed wave equations obtained in this
way can be linearized. One obtains for the symmetrized and irreducible
tensor field of integer spin:
\beq
P^I g_{IJ} A^{J \cdots K}({\cal P}) = 0
\eeq
For a state of half integral spin in a general Rarita-Schwinger basis in which
the fields are irreducible with respect to the QLGr one finds:
\beq
\beqcol
P^I g_{IJ} \chi^{J\cdots K\,\tde{\alf}}({\cal P})  & = & 0,  \\
(P_{{\xi}_{(1/2)}})\indf{\beta}{\tde{\alf}}  \chi^{\tde{\alf}\, J\cdots
K}({\cal P}) &=&
\psi^{\beta\, J}({\cal P})
\eeqcol \qquad
\beqcol
P^I g_{IJ} \psi^{J \cdots K \, \beta}({\cal P}) & = &  0 ,     \\
(P_{{\bar{\xi}}_{(1/2)}})\indf{\tde{\alf}}{\beta} \psi^{\beta\, J\cdots
K}({\cal P}) &=&
\chi^{\tde{\alf}\, J}({\cal P}).
\eeqcol
\eeq
It should also be mentioned
that the procedure outlined in section 6.2 for obtaining
massless wave equations from the massive ones applies directly to
higher dimensional systems. These equations take the form e.g.:
\beq
\Pxi{\half}\indf{\alf}{\tde{\beta}}
\phi^{\tde{\beta}\cdots\tde{\gamma}}({\cal P}_{m^2=0}) = 0
\eeq

\section{Summary}

\bigskip

In the sense of Wigner \cite{WIG} elementary particles are considered as
unitary Irreps of the Poincar\'e group. The $q$-deformation of the
Poincar\'e symmetry leads in a first step \cite{PSW} to spinless
particles which live on a non-commutative momentum Minkowski space. This
has the effect that the spectra of energy, the measurable component
of the space-momenta and the mass admit a discretization.
In a second step one goes over to spinor states \cite{PW} of the
$q$-deformed Poincar\'e symmetry which are not a unitray Irrep. These
states have more spin degrees of freedom than the physical particle has.
The $q$-deformed wave equations remove the additional degrees of
freedom. This proves that the unitary Irreps of the $q$-deformed
Poincar\'e symmetry actually are classified by the stability subgroup
$\suq$. One has therefore a representation therory analogous to the undeformed
case. The analysis of this work solves the $q$-deformed relativistic
one-particle problem in a non-commutative momentum space from the viewpoint
of representation theory.

\section{Appendix}

\smallskip

{\bf a.} {\sl The $\eps$ - tensors}

The tensors which are used for raising and lowering $\slq$ spinor indices
are \cite{CSSW}:
\goodbreak
\beq
\eps_{\alf\beta}=\mattwo{0}{q^{-\half}}{-q^{\half}}{0}_{\alf\beta}, \qquad
\eps^{\alf\beta}=\mattwo{0}{-q^{-\half}}{q^{\half}}{0}^{\alf\beta} ,
\label{eps}
\eeq
with normalization $\eps_{\alf\beta}\eps^{\beta\gamma}=
\delta^{\gamma}_{\alf}$. The $\eps$'s with tilded indices are identical to
those in (\ref{eps}).

\medskip

{\bf b.} {\sl $q$-numbers}

A $q$-number is defined by:
\beq
[n]_{q^{a}} = {{1- q^{a\, n}}\over {1 - q^a}}
\label{qnum}
\eeq
The $q$-deformed binomial coefficients are given by the definition:
\beq
\left[ \begin{array}{c}
         n \\ k       \end{array} \right]_{q^a}  = {{[n]_{q^a}!}\over{
[k]_{q^a}! [n-k]_{q^a} !}}
\label{qbinom}
\eeq

\medskip

{\bf c.} {\sl Boost matrices in an arbitrary representation}

The higher dimensional representations of the boost matrices $w_B$
introduced in section 3 are special cases of general $\slq$ matrices whose
explicit form can be found in \cite{MAS}. Since the $(w^{(l)}_B)$'s are
used in the construction of general $q$-deformed wave equations they are
listed here.

Case I: $ i+j \leq 0, \quad j \geq i$:
\beq
(w_B^{(l)})^i_j = \rho^{-i-j} z^{j-i} \binom{l-i}{j-i}\binom{l+j}{j-i}
                q^{(l+i)(i-j)}
\label{w1}
\eeq
Case II: $ i+j \geq 0, \quad j \geq i $:
\beq
(w_B^{(l)})^i_j = z^{j-i} \rho^{-i-j}\binom{l-i}{j-i}\binom{l+j}{j-i}
                  q^{-(j-i)(j-l)}
\label{w2}
\eeq
The complex conjugate representations are obtained by $\bar{w}^{(l)}_B =
k(w^{(l)}_B)$ using the mapping $k$ of section 3 and
$ j\left((w^{(l)}_B)\indf{\alf}{\beta} \right) = (-q)^{j-i}
           (\bar{w}^{(l)}_B)\indf{-\tde{\alf}}{-\tde{\beta}} $. This has
been shown in \cite{PW}.

\medskip

{\bf d.} {\sl Explicit form of the $q$-Rarita-Schwinger operator ${\cal R}_1$}

\begin{eqnarray}
{\cal R}_1{}^{\tde{\alf}\tde{\beta}\gamma}{}_{\tde{\mu}\nu\rho} &=&
{\hat R}^{\tde{\alf}\tde{\beta}}{}_{\tde{\beta}_1\tde{\alf}_1}
{\hat R}^{-1}{}^{\tde{\alf}_1\gamma}{}_{\gamma_1\tde{\alf}_2}
{\hat R}^{-1}{}^{\tde{\mu}_1\tde{\alf}_2}{}_{\tde{\alf}_3\tde{\mu}}
\eps_{\nu\nu_1}{\hat R}^{-1}{}^{\tde{\alf}_3\nu_1}{}_{\nu_2\tde{\alf}_4}
\eps^{\nu_2\nu_3}
{\hat R}^{-1}{}^{\tde{\beta}_1\gamma_1}{}_{\gamma_2\tde{\beta}_2}
{\hat R}^{-1}{}^{\tde{\mu}_2\tde{\beta}_2}{}_{\tde{\beta}_3\tde{\mu}_1}
\times \nonumber \\
& \; & \qquad
\times (P_{\xi})\indf{\gamma_2}{\tde{\mu}_2}
(P_{\bar{\xi}})\indf{\tde{\beta}_3}{\nu_3}
(P_{\bar{\xi}})\indf{\tde{\alf}_4}{\rho}
\label{rsexp}
\end{eqnarray}

\bigskip
\bigskip

{\bf Acknowledgement}

The author would like to thank Professor J. Wess for drawing his attention
to the problem of $q$-deforming relativistic wave equations and
A. Schirrmacher, J. Schwenk, M. Niedermaier  and W.B. Schmidke for
stimulating and helpful discussions.

%
%
%
%
%
%

\end{document}